\documentclass[showpacs,preprintnumbers,amsmath,amssymb]{revtex4}
\usepackage{graphicx}
\usepackage{dcolumn}
\usepackage{xcolor}
\usepackage{bm}
\begin{document}
\title{Radiative transition of an atom falling into spherically
symmetric Lorentz violating black hole background}
\author{Anisur Rahaman}
\email{manisurn@gmail.com} \affiliation{Durgapur Government
College, Durgapur-713214,  India}
\date{\today}
\begin{abstract}
In this work, we explore the intriguing phenomenon of acceleration
radiation exhibited by an atom falling into a black hole, as
previously studied in Phys. Rev. Lett. 121, 071301 (2018) . Our investigation focuses on
examining the impact of Lorentz violation within the framework of
the bumblebee gravity model on this phenomenon. We observe that the excitation probability although acquires
Planck-like factor the exponential part of it acquires the Lorentz
violation factor dependent frequency. However, equivalence principle is not  violated. Then we calculate
the horizon brightened acceleration radiation (HBAR) entropy for
this black hole geometry. We observed that the HBAR entropy has
the form similar to that of Bekenstein-Hawking black hole entropy
however it has been observed that it is also influenced by Lorentz
violation associated to the  Bumblebee theory. Additionally, we note that the Lorentz violation effect and conformal symmetry both affect the transition probabilities of a two-level atomic detector.
\end{abstract}
\maketitle
 \section{Introduction}
 The fundamentals of general relativity, as first proposed by Einstein
 \cite{EIN15, EIN16}, form the basis for the unification of geometry and gravity, giving rise to the concept of black holes as precise solutions to the non-linear field equations of the theory. Fifty years after its development, the integration of general
relativity with thermodynamics led to several remarkable
discoveries. In his seminal works, Stephen Hawking \cite{HAW1,
HAW2} demonstrated that black holes emit radiation when quantum
effects in curved spacetime are considered. This gave birth to
fields such as black hole thermodynamics \cite{HAW1, HAW2, BAKE1,
BAKE2}, Hawking radiation \cite{HAW1, HAW2, BAKE1}, and studies on
 particle emission from black holes \cite{PAGE1, PAGE2, PAGE3}, along with important insights into the Unruh effect \cite{UNRH} and acceleration radiation \cite{DEWITT, UNRUH2, MULLER, Vanzella, Higuichi,  ONRDENZ1, ONRDENZ2, ONRDENZ3, ONRDENZ4, RITU}.

One of the most intriguing areas of recent research is the study
of the thermal and geometrical properties of black hole horizons
and their close connection to the dynamics of particles near these regions. 
The detection of particles by an accelerated observer in an inertial vacuum, also known as the Unruh effect \cite{UNRH}, can be directly linked to Hawking radiation. The interplay between black hole physics and quantum optics have gained significant attention in recent years, with efforts focused on providing analytical explanations of these combined effects \cite{DEWITT, UNRUH2, MULLER, Vanzella, Higuichi, ONRDENZ1, ONRDENZ2, ONRDENZ3, ONRDENZ4, RITU}. In this context, the two-level atom acts as a
detector of these phenomena. Recent studies have demonstrated that atoms falling into a black hole experience thermal radiation
effects similar to Hawking radiation, regardless of the underlying spacetime metric explicitly admits a black hole solution \cite{PAGE0, RITU, WEISS, PHILBIN}. The concept of
Horizon Brightened Acceleration Radiation Entropy (HBAR) describes
the radiation entropy emitted during this process, providing a
useful framework for understanding the connection between general
relativity and atom optics \cite{PAGE0}.

Moreover, some studies have investigated how conformal symmetry
influences the analysis of black hole near-horizon features
\cite{CONF1, CONF2, ONRDENZ1, ONRDENZ2, ONRDENZ3, ONRDENZ4}.
These works focus on the behavior of near-horizon elements shedding
light on the symmetry properties of these regions. On the other
hand, quantum mechanics remains a fundamental pillar of
theoretical physics, and its unification with general relativity
has proven to be one of the greatest challenges in the field. The quest
for a comprehensive theory of quantum gravity has been a central focus
of theoretical research \cite{ROVELLI, CARLIP, ACV,
KOPAPR}, but despite considerable efforts, a complete theory
remains elusive. As a result, various alternative approaches to
quantum gravity have gained attention in recent years. Bumblebee
theory \cite{BUMBLEBEE, BUMBLES1,BUMBLES2,BUMBLES3, BUMBLEBEE1} is remarkable 
in this context where Lorentz symmetry breaks
spontaneously when the bumblebee field acquires non-vanishing
vacuum expectation value with consequent violation of $U(1)$
symmetry. This model belongs to the standard model extension.

An intriguing theoretical topic is the acceleration radiation
phenomena that an atom falling into a black hole exhibits. This
phenomenon has been examined in several black hole backgrounds,
and the impact of quantum correction  has also been investigated.
A theory known as bumblebee gravity incorporates the quantum
effect through an indirect impact that promotes spontaneous
Lorentz violation. Several attempts have been made to study the
Lorentz violation in different contexts. Therefore, a study of the
effect of Lorentz violation on the acceleration radiation
phenomenon displayed by an atom falling into a black hole using
the bumblebee theory is interesting and instructive.

The equivalence principle stands as a fundamental cornerstone of general relativity, and any departure from it is, in principle, undesirable. Nevertheless, instances suggesting its violation have long fueled intriguing debates. Discussions on the possible breakdown of the equivalence principle trace back decades and continue to this day \cite{EQP0, EQP1, EVS0, EVS1, EVPAGE, EVS2, EVS3, EQP2, EQP3, EQP4, EQP5}, largely due to the lack of a conclusive resolution. Recent studies indicate that this principle can indeed be violated, as revealed through the response function of an Unruh–DeWitt detector across different spacetime geometries and vacuum states \cite{SINGLETON, SINGLETON1}. Explicit examples of such violations have been demonstrated in \cite{VEQUIV}, where quantum corrections play a decisive role. This naturally raises a compelling question: Are quantum corrections at the root of these anomalous effects? In particular, the metric under consideration acquires a Lorentz-violating term via indirect quantum effects. This makes the fate of the equivalence principle in such scenarios especially intriguing, warranting a careful and systematic investigation. Of particular interest is the deeper question: How does Lorentz symmetry violation influence the validity of the equivalence principle? Exploring this intersection not only sharpens our understanding of these anomalies but also promises to shed light on the very foundations of modern physics

The paper is organized as follows. Section II provides a brief overview of the Lorentz-violating bumblebee model. In Section III, we present the equivalent Rindler form of this Lorentz-violating background. Section IV is devoted to solving the Klein–Gordon field equation in this background. In Section V, we compute the emission and absorption probabilities and examine the influence of Lorentz violation on these processes. Section VI is concerned with the computation of the modified HBAR entropy. Finally, Section VII contains a summary of the main results and concluding remarks.

\section{Lorentz violating bumblebee background in brief:
Bumblebee Model} In the bumblebee gravity  \cite{BB0}  model
Lorentz symmetry breaking is manifested through a bumblebee field
vector $B_\mu$. This model is  described by the action
\begin{equation}
S_{BB}=\int d^4x[\frac{1}{16\pi G_N}({\cal R}+ \upsilon
B^{\mu}B^{\nu}{\cal R}_{\mu\nu})-\frac{1}{4}{\cal B}_{\mu\nu}{\cal
B}^{\mu\nu}-V(B_\mu B^\mu \pm b^2)]. \label{BA}
\end{equation}
Here $\upsilon$ is a dimension-full parameter having the dimension of
$M^{-2}$.  The field strength tensor associated with the bumblebee
field $B_\mu$ is defined by
\begin{equation}
{\cal B}_{\mu\nu} = \nabla_\mu B_\nu - \nabla_\nu B_\mu,
\end{equation}
and $R_{\mu\nu}$ stands for Ricci tensor. The potential $V$ defined in
terms of the bumblebee field is at the root of the breaking  of
Lorentz symmetry. When the bumblebee field collapses onto a
nonzero minima maintaining  condition $B_\mu B^\mu = \pm b^2$. The
Lagrangian density of the bumblebee gravity model \cite{BB0, BB1,
BB2} results from the following extended  Einstein equations
\begin{eqnarray}
G_{\mu\nu}=R_{\mu\nu}-\frac{1}{2}Rg_{\mu\nu}=\kappa
T_{\mu\nu}^{B}.
\end{eqnarray}
Here  $G_{\mu\nu}$ and $T_{\mu\nu}^{B}$  are respectively
representing Einstein and bumblebee energy-momentum tensors. The
gravitational coupling is defined by ${\kappa=8\pi G}_{{N}}$, and
the bumblebee energy-momentum tensor $T_{\mu\nu}^{B}$ are provided
by the expression
\begin{eqnarray}
T_{\mu\nu}^{B} & =
&-B_{\mu\alpha}B_{~\nu}^{\alpha}-\frac{1}{4}B_{\alpha\beta
}B^{\alpha\beta}g_{\mu\nu}-Vg_{\mu\nu}+2V^{\prime}B_{\mu}B_{\nu}+\frac{\upsilon
}{\kappa}\left[
\frac{1}{2}B^{\alpha}B^{\beta}R_{\alpha\beta}g_{\mu\nu
}-B_{\mu}B^{\alpha}R_{\alpha\nu}\right.\nonumber  \\
& - & B_{\nu}B^{\alpha}R_{\alpha\mu}+\frac{1}{2}\nabla_{\alpha}
\nabla_{\mu}(  B^{\alpha}B_{\nu})  +\frac{1}{2}\nabla_{\alpha
}\nabla_{\nu}(  B^{\alpha}B_{\mu})  -\frac{1}{2}\nabla^{2}(
B_{\mu}B_{\nu}) -\frac{1}{2}g_{\mu\nu}\nabla_{\alpha}\nabla_{\beta
}( B^{\alpha}B^{\beta})].
\end{eqnarray}
The parameter $\upsilon$ having dimension $M^{-1}$ is representing
the real coupling constant in this situation. Prime denotes the derivative with respect to the argument.
Prime denotes the derivative with respect to the argument.
Therefore,
\begin{equation}
V'= \frac{\partial V(z)}{\partial z}|_{z=B_\mu B^\mu\pm b^2}.
\end{equation}
The equation of motion of bumblebee field is  provided by
\begin{equation}
\nabla^{\mu}{\cal B}_{\mu\nu}= 2V'B_\nu
-\frac{\upsilon}{\kappa}B^\mu, R_{\mu\nu}.
\end{equation}
and the covariant divergence of the bumblebee energy-momentum
tensor reads
\begin{equation}
\nabla^{\mu}T_{\mu\nu}^{B}=0. \label{CON}
\end{equation}
which is  the  covariant conservation law for the energy-momentum
tensor Ultimately, we have
\begin{equation}
R_{\mu\nu}=\kappa
T_{\mu\nu}^{B}+\frac{\upsilon}{4}g_{\mu\nu}\nabla^{2}\left(
B_{\alpha}B^{\alpha}\right)  +\frac{\upsilon}{2}g_{\mu\nu}\nabla_{\alpha}%
\nabla_{\beta}(B^{\alpha}B^{\beta}).
\end{equation}
It is transparent that we recover the ordinary Einstein equations
when the bumblebee field $B_{\mu}$ vanishes. When the bumblebee
field acquires vacuum expectation value (VEV) we are allowed to
write
\begin{equation}
B_{\mu}=b_{\mu},
\end{equation}
Consequently,
\begin{equation}
{\cal B}_{\mu\nu}\equiv\partial_{\mu}b_{\nu}-\partial_{\nu}
b_{\mu}.
\end{equation}
In this situation  particular form of the potential driving the
dynamics is irrelevant. And consequently, we have $V = 0, V' = 0$
and the Einstein equation acquires a generalized expression
\begin{eqnarray}
& & R_{\mu\nu}+\kappa
b_{\mu\alpha}b_{~\,\nu}^{\alpha}+\frac{\kappa}
{4}b_{\alpha\beta}b^{\alpha\beta}g_{\mu\nu}+\upsilon
b_{\mu}b^{\alpha}R_{\alpha\nu }+\upsilon
b_{\nu}b^{\alpha}R_{\alpha\mu}-\frac{\upsilon}{2}b^{\alpha}b^{\beta}
R_{\alpha\beta}g_{\mu\nu}\nonumber\\
&-&\frac{\upsilon}{2}\nabla_{\alpha}\nabla_{\mu}(
b^{\alpha}b_{\nu}) -\frac{\upsilon}{2}\nabla_{\alpha}\nabla_{\nu}(
b^{\alpha}b_{\mu}) +\frac{\upsilon}{2}\nabla^{2}\left(
b_{\mu}b_{\nu}\right)  = 0. \label{GEE}
\end{eqnarray}
The space-like background for the bumblebee field $b_{\mu}$ allows
to  choose $b_{\mu}=[0,b_{r}(r),0,0]$. Consequently, we have
$b^{\mu}b_{\mu}=b^{2}=$\textit{constant} and thus the  spherically
symmetric static vacuum solution to Eqn. (\ref{GEE}) emerges
as \cite{BB1}
\begin{equation}
ds^{2}=-\left(  1-\frac{2\mathcal{M}}{r}\right)  dt^{2}+\left(
1+l\right) \left( 1-\frac{2\mathcal{M}}{r}\right)
^{-1}dr^{2}+r^{2}d\Omega^2, \label{SPTS}
\end{equation}
where $d\Omega^2=\left( d\theta^{2}+\sin^{2}\theta
d\varphi^{2}\right)$, and  the parameter $\ell$ is  termed as
Lorentz symmetry breaking (LSB) parameter and it is precisely
given by $\ell=\upsilon b^{2}\geq 0$. Kretschmann scalar for
this metric reads
\begin{equation} \mathcal{K}=\frac{4\left(
12\mathcal{M}^{2}+4\ell\mathcal{M}r+\ell^{2}r^{2}\right)
}{r^{6}\left( 1+\ell\right) ^{2}}. \label{KSCALAR}
\end{equation}
As this metric renders the Kretschmann   scalar (\ref{KSCALAR})
 is different from the Schwarzschild metric \cite{BB1}, the
metric is known to be very different although it has structural
similarity with the Schwarzschild metric.  The Hawking temperature
corresponding to the metric is provided by
\begin{equation}
T_{H}=\frac{1}{4\pi\sqrt{-g_{tt}g_{rr}}}\left.
\frac{dg_{tt}}{dr}\right\vert _{r=r_{h}}=\left.
\frac{1}{2\pi\sqrt{1+l}}\frac{\mathcal{M}}{r^{2}}\right\vert
_{r=r_{h}}=\frac{1}{8\pi \mathcal{M}\sqrt{1+\ell}}. \label{HATEM}
\end{equation}
Eqn. (\ref{HATEM}) as a consequence, acquires a precise $\ell$
dependent generalized expression. The Kerr-like Bumblebee solution
is developed in \cite{BB20, BB2} and there are several application
on bumblebee black hole solution in diverse direction of
astrophysics and cosmology \cite{BB1,  BB20, BB2, BB3, BB4, BB5,
BB6}. These are all known about the bumblebee gravity. Therefore,
the extent to which the classical dynamics in bumblebee gravity
differ from the Schwarzschild would be instructive.

According to the solution (\ref{SPTS}) the line element in the
Lorentz violating Schwarzschild background given by
\begin{equation}
ds^2=-f(r)c^2dt^2 + g(r)dr^2 + r^2d\Omega^2. 
\label{LVSCH}
\end{equation}
where
\begin{equation}
f(r)=1-\frac{2\mathcal{M}G}{rc^2},  g(r)=(1+\ell)f(r)^{-1},
\label{HORE}
\end{equation}
and, $G$ is the gravitational constant $\mathcal{M}$ is the mass
of the black hole and $c$ is the velocity of light. Eqn.
(\ref{HORE}) was in natural unit where $G$ and $c$ were taken
unity.  Our plan for this work is to consider atoms falling into
the black holes associated with the bumblebee gravity black hole
which contains Lorentz violation and draw conclusions from such a
thought experiment.

An important aspect of this type of work is to get an insight into
the Einstein's principle of equivalence. There were several
attempts to provide an alternative explanation of the equivalence
principle \cite{Rohrlich}-\cite{SG}. Such an insight was obtained
earlier in \cite{FULLING, SG} where the spontaneous excitation of
a two-level atom was studied in the presence of a perfectly reflecting mirror
in the generalized uncertainty principle framework. It
was shown in \cite{SG} that when the mirror is accelerating, the
 the probability of excitation of the atom
gets modulated due to the generalized uncertainty principle,
thereby leading to an explicit violation of the equivalence
principle. In the present context, an atom falling into a black hole is predicted to emit radiation characterized by a thermal spectrum, analogous to that produced by a stationary atom in the presence of an accelerating mirror. Since quantum corrections play a role in this phenomenon, it becomes a matter of deep interest to examine whether the fundamental cornerstone of general relativity—the equivalence principle—remains valid under such circumstances. The second aim is to lookfor quantum gravity corrections in the HBAR entropy and see
whether they are logarithmic in nature similar to the corrections in the Bekenstein-Hawking entropy \cite{BAKE1, BAKE2, HAW1, HAW2,
PK}.

\section{The Lorentz violating background in Rindler form}
Let us begin with the  Minkowski metric in (1+1)dimensional space
time
\begin{equation}\label{MIN}
ds^2=c^2dt^2-dx^2.
\end{equation}
If a body moves  with a uniform proper acceleration `$a$' in this
$(1+1)$ dimensional flat spacetime, in terms of the proper time
$\tau$ the position and time coordinates  can be expressed
respectively by
\begin{align}
x(\tau)&=\frac{c^2}{a}\cosh(\frac{a\tau}{c}),\label{T1}
\end{align}
\begin{align}
t(\tau)&=\frac{c}{a}\sinh(\frac{a\tau}{c}).\label{T2}
\end{align}
The  Eqn.(\ref{MIN}), is satisfied  by the definition of position
and time given in equations (\ref{T1}, \ref{T2}). We bring in
action the following  coordinate transformations:
\begin{align}
x&=\eta\cosh(\frac{\tilde{a}\tilde{t}}{c}),\label{TT1}
\end{align}
\begin{align}
t&=\frac{\eta}{c}\sinh(\frac{\tilde{a}\tilde{t}}{c}).\label{TT2}\\
\end{align}
Under the transformation  (\ref{TT1}) and (\ref{TT2}) the metric
(\ref{MIN}) gets modified to Rindler form
\begin{equation}
ds^2=\left(\frac{\tilde{a}\eta}{c^2}\right)^2c^2d\tilde{t}^2-d\eta^2.\label{RMET}
\end{equation}
The Rindler form of the metric  standing in equation (\ref{RMET})
describes a uniform acceleration of a body. A  comparison between
the Eqn.(\ref{T1}) and Eqn.(\ref{TT1}), reveals that the proper
time of the moving body can be written down as
\begin{equation}
\tau=\frac{\tilde{a}\tilde{t}}{a}\label{PRT}
\end{equation}
Ultimately we get  the  expression of the uniform acceleration
of the body in Minkowski spacetime comparing Eqn.(\ref{T2}) and
Eq.(\ref{TT2}):
\begin{equation}
a=\frac{c^2}{\eta}. \label{UNIAC}
\end{equation}
With this mathematical input now, make an attempt to reduce the
Lorentz-violating Schwarzschild spacetime in the Rindler form
that enables us to have  the uniform acceleration of a freely
falling body in the vicinity of the event horizon. The event
horizons of the Lorentz-violating Schwarzschild black hole can be
obtained by  setting $f(r)=0$ in Eqn.(\ref{HORE}). The position 
of theevent the horizon remains the same as the Schwarzschild black hole
i.e. at $r_{eh}= 2M$ in spite of the amendment of the quantum
correction due to the Lorentz-violating effect caused by the
bumblebee field  entered in to the picture.

We will now effectuate a near horizon expansion with respect
to the event horizon to express the Lorentz-violating
Schwarzschild metric in the Rindler form.  A Taylor series
expansion of $f(r)$ about the event horizon $r_{eh}$  will serve
this purpose. The near horizon expansion is executed keeping terms
up to first order in the near horizon expansion parameter
$(r-r_{eh})$. So it results
\begin{equation}
f(r)\cong
f(r_{r_{eh}})+(r-r_{eh})\frac{df(r)}{dr}\biggr|_{r=r_{eh}}=(r-r_{eh})f'(r_{eh}).
\label{EXPHOR}
\end{equation}
This expansion enables us to obtain the line element in $(1+1)$
dimensions spacetime as follows:
\begin{equation}
\begin{split}
ds^2&=f(r)c^2dt^2-f(r)^{-1}dr^2\\
&\cong(r-r_{eh})f'(r_{eh})c^2dt^2-\frac{1}{(r-r_{eh})f'(r_{eh})}dr^2.
\label{APPDS}
\end{split}
\end{equation}
To express (\ref{APPDS}) into the Rindler form we now define a new
coordinate $\eta$, which is given by

\begin{equation}
\eta=2\sqrt{\frac{(r-r_{eh})(\ell+1)}{f'(r_{eh})}}.\label{ETA}
\end{equation}
With the the above definition Eqn.(\ref{ETA}), we land on to
Rindler form of the metric in $(1+1)$ dimensional spacetime:
\begin{equation}
ds^2\cong \frac{\eta^2f'^2(r_{eh})}{4}c^2dt^2-d\eta^2.
\label{RINDSCH}
\end{equation}
Here $f'(r_{eh})$ reads
\begin{equation}
f'(r_{eh})=\frac{c^2}{2MG}\label{ORF},
\end{equation}
If we retainin only the first order term in Taylor's series
Comparing Eqn.(\ref{RINDSCH}) with Eqn.(\ref{UNIAC}), we obtain
uniform acceleration corresponding to curves of constant $\eta$:
\begin{equation}
\begin{split}
a&=\frac{c^2}{\eta}=\frac{c^2}{2\sqrt{\frac{r-r_{eh}}{f'(r_{eh})}}}\\
&\cong\frac{c^2\sqrt{f'(r_{eh})}}{2\sqrt{r}}\left(1+\frac{r_{eh}}{2r}\right)
\end{split}.\label{DEFAC}
\end{equation}
\section{Solution of Klein Gordon field equation in the Lorentz violating background}
We consider the spherically symmetric bumblebee gravity inspired
Lorentz violating geometry in $D$ spacetime dimensions. The metric
of which  is given by
\begin{equation}
ds^2=-f(r)c^2dt^2+ g(r)dr^2+r^2d\Omega^2_{(D-2)}.
\label{LVSCHD}
\end{equation}
Here, $f(r)= 1-\frac{2\mathcal{M}}{r^{D-3}}$, and $g(r)=(1+\ell)f(r)^{-1}$. If we take into account
the minimal coupling of this background  with scalar field $\Phi$
the action in that case will be given by
\begin{equation}
{\cal A}= -\frac{1}{2}\int d^D x
\sqrt{-g}[g^{\mu\nu}\partial_\mu\Phi\partial_\nu\Phi+ m_0^2\Phi].
\label{ACT}
\end{equation}
The  Klein-Gordon type  equation for the scalar field that follows
from Eqn. (\ref{ACT}) reads
\begin{equation}
\frac{1}{\sqrt{-g}}\partial_\mu(\sqrt{-g}g^{\mu\nu}\partial_\nu\Phi)-m_0^2\Phi=0.
\label{KGE}
\end{equation}
The most general solution of the scalar $\Phi$ satisfying equation
(\ref{KGE}) would be of the form
\begin{equation}
\Phi(t,r, \Omega )=\sum\limits_{n,l,m}\left[a_{nlm}\phi_{nlm}(t,r,
\Omega )+ H.C.\right], \label{PHI}
\end{equation}
where $a_{nlm}$ stands for  the annihilation operator of the field
under consideration, and $a^{\dagger}_{nlm}$ is the corresponding
creation operator. To have a solution of the Eqn. (\ref{KGE})
we employ standard separation of variables prescription for
$\phi_{nlm}$:
\begin{equation}
\Phi_{nlm}=\xi
r)u_{nl}(r)Y_{lm}(\Omega)e^{-i\nu_{nl}t}.\label{SPV}
\end{equation}
Here $Y_{lm}(\Omega)$  is the spherical harmonics and  $\nu_{nl}$
is denoting the frequency of the scalar field. Keep in view that
$\Phi_{nlm}$ satisfies the traditional Klein-Gordon field
equation, we adopt the following version of $\eta(r)$:
\begin{eqnarray}
\xi(r)&=&\exp[-\frac{1}{4}\int dr(\frac{f'
(r)}{f(r)} + \frac{g'(r)}{g(r)}+\frac{2(D-2)}{r})]\nonumber \\
 &=&(f(r)g(r))^{-\frac{1}{4}}r^{-(\frac{D-2)}{2}}\nonumber\\
 &=&(\ell+1)^{-\frac{1}{2}}r^{-(\frac{D-2}{2})}. \label{XI}
\end{eqnarray}
If we we write Eqn.(\ref{KGE}) explicitly in terms of temporal,
radial, and angular coordinates we have
\begin{equation}
\begin{split}
&-\frac{\ddot\Phi}{f(r)}+\left(\frac{D-2}{r}+\frac{f'(r)}{f(r)}\right)\frac{f(r)}{\ell+1}\Phi'
+\frac{f(r)}{\ell+1}\Phi''+\frac{1}{r^2}\Delta_{D-2}\Phi-m_0^2\Phi=0
\end{split}
\label{RT}
\end{equation}
where $\Delta_{D-2}$ is the,  Laplacian on the unit
$(D-2)$-dimensional sphere ($S^{D-2}$). To have a solution of the
field in this curved background, we plug in Eqn.(\ref{SPV}) and
Eqn. (\ref{XI}) in Eqn.(\ref{RT}), and  land on to the following
radial equation
\begin{equation}
\begin{split}
&u_{nl}''(r)+\biggr[-\left(\frac{f''(r)}{2f(r)}+\frac{D-2}{2r}\left(\frac{f'(r)}{f(r)}\right)\right)
+\left(\frac{(\ell+1)\nu_{nl}^2}{(f(r))^2}+\frac{1}{4}\left(\frac{(f'(r))^2}{(f(r))^2}\right)\right)\\
&-\frac{(\ell_D+m^2_0r^2)(\ell+1)}{r^2f(r)}+\biggr(\left(\frac{(\ell+1)-f(r)}{f(r)}-1\right)\frac{(D-3)^2}{4r^2}
+\frac{1}{4r^2}\biggr)\biggr]u_{nl}(r)=0,
\end{split}
\label{UEQ}
\end{equation}
where $\ell_D=\left(l+\frac{D-3}{2}\right)^2$. Let us now 
execute our analysis by extending it away from the immediate
horizon, i.e. beyond near horizon approximation. The form of
$f(r)$ and its higher order derivatives in this beyond near
horizon prescription can be written down as follows.
\begin{align}
f(r)&\cong (r-r_{eh})f'(r_{eh})+\frac{(r-r_{eh})^2}{2}f''(r_{eh})\nonumber\\
&+\mathcal{O}((r-r_{eh})^3),\label{FRA}\\
f'(r)&\sim f'(r_{eh})+(r-r_{eh})f''(r_{eh})~,\label{FRB}\\
f''(r)&\sim f''(r_{eh})\label{FRC}~.
\end{align}
For notational convenience we will use $r-r_{eh}=z$,
$u_{nl}(r)=u$, $\nu_{nl}=\nu$ through out the rest part of the
body of this paper. In this beyond near-horizon approximation
Eqn.(\ref{UEQ}) can be casted as follows
\begin{equation}
\begin{split}
&u''+\left(\frac{\nu^2(\ell+1)}{(f'(r_{eh})^2}+\frac{1}{4}\right)\frac{u}{z^2}
-\biggr[\left(\frac{\nu^2(\ell+1)}{(f'(r_{eh})^2}+\frac{1}{4}\right)\left(\frac{f''(r_{eh}}{f'(r_{eh})}
\right)
\\&+\frac{(D-2)}{2r_{eh}}+\frac{(\ell_D+r_{eh}^2m_0^2)(\ell+1))}{(r_{eh})^2}
-\frac{(\ell+1))(D-3)^2}{4f'(r_{eh})(r_{eh})^2}\biggr]\frac{u}{z}=0~.
\end{split}
\label{UEQC}
\end{equation}
A more intelligible manner of expressing  Eqn.(\ref{UEQC}) can be
performed:
\begin{equation}
z^2u''-Q z
u+\left(\frac{(\ell+1)\nu^2}{(f'(r_{eh}))^2}+\frac{1}{4}\right)u=0,
\label{UEQS}
\end{equation}
where  $\kappa$ has the expression
\begin{equation}
\begin{split}
Q&=\left(\frac{(\ell+1)\nu^2}{(f'(r_{eh}))^2}+\frac{1}{4}\right)\left(\frac{
f''(r_{eh})}{
f'(r_{eh})}\right)+\frac{(D-2)}{2r_{eh}}\\&+\frac{(\ell+1)(\ell_D+r_{eh}^2m_0^2)}{(r_{eh})^2f'(r_{eh})}
-\frac{(\ell+1)(D-3)^2}{4f'(r_{eh})(r_{eh})^2}~.
\end{split}
\label{KAPPA}
\end{equation}
We  are now in a position  to proceed towards obtaining the
solution of the radial equation.  Since $r-r_+=z$ renders very
small contribution, the term
$\mathcal{O}\left(\frac{1}{z^2}\right)$ provides the most dominant
contribution in Eqn.(\ref{UEQC}). If we neglect the contribution
of the term $\mathcal{O}\left(\frac{1}{z}\right)$ of
Eqn.(\ref{UEQC}), we  land on to
\begin{equation}
z^2u''(z)+\left(\frac{(\ell+1)\nu^2}{(f'(r_{eh}))^2}+\frac{1}{4}\right)u(z)=0~.
\label{APPE}
\end{equation}
There is something intriguing about the equation \ref{APPE}: a
careful look reveals that an asymptotic conformal symmetry is
present within this. By looking at the following ansatz, this may
be seen:
\begin{equation}
u(z)=z^s~. \label{ANSAT}
\end{equation}
Under substitution of this ansatz  in Eqn.(\ref{APPE}), a
quadratic equation in $s$  results:
\begin{equation}
s^2-s + \frac{(\ell+1)\nu^2}{(f'(r_{eh}))^2}+\frac{1}{4}=0~.
\label{EQS}
\end{equation}
After resolving the preceding equation, we can find an analytical
expression of  $s$:
\begin{equation}
s=\frac{1}{2}\pm i\frac{\nu\sqrt{\ell+1}}{f'(r_{eh})}~.
\end{equation}
It is possible to recast the solution of equation (\ref{APPE}) in
the  following comprehensive form
\begin{equation}
u(z)=z^{\frac{1}{2}\pm i\frac{\nu\sqrt{\ell+1}}{f'(r_{eh})}}~.
\label{UZE}
\end{equation}
Note that the solution of the  equation (\ref{UZE}) has both ingoing
and outgoing component. We are interested here in the out going
components of it. The equation (\ref{UZE}) provides
\begin{equation}\label{1.22}
u(z)=z^{\frac{1}{2}+ i\frac{\nu\sqrt{\ell+1}}{f'(r_{eh})}},
\end{equation}
as the out going component. Additionally, scaling solutions for
Eqn. (\ref{APPE}) are permitted, which confirms the conformal
symmetry of the problem. Moreover, we can see that the near
horizon physics is equivalent to a one-dimensional effective
Hamiltonian
\begin{equation}\label{1.22a}
{\cal H}=P_z^2-\frac{\zeta}{z^2}.
\end{equation}
where $P_z= -i\frac{\partial}{\partial z}$ in natural unit ($\hbar=1$),
$\zeta=\left(\frac{(\ell+1))\nu^2}{(f'(r_{eh}))^2}+\frac{1}{4}\right)>0$.
This indeed depicts the widely recognised long-range conformal
quantum mechanics \cite{QMA}. Consequently, the position and time
dependent component of the solution is provided by
\begin{equation}
\phi(r,t)=e^{-i\nu t}\zeta(r) u(r)={\cal C}e^{-i\nu
t}z^{i\frac{\nu\sqrt{\ell+1}}{f'(r_{eh})}},\label{PHII}
\end{equation}
with
$\mathcal{C}=(r_{eh})^{\frac{D-2}{2}}(\frac{f'(r_{eh})}{\sqrt{\ell+1}})^{-\frac{1}{4}}$.
We will now examine the scenario in which Eqn.(\ref{EQS}) now
includes the $\mathcal{O}(1/z)$ term. Since Eqn.(\ref{UEQS}) lacks
conformal symmetry, the ansatz in Eqn.(\ref{ANSAT}) will not
succeed at this time. We consider a solution in the manner
specified by Eqn. (\ref{PHI}), and assume that $u(z)$ has the following series expansion:
\begin{equation}
u(z)=\sum\limits_{k=0}^\infty A_k z^{k+ p}. \label{UZES}
\end{equation}
where a constant $p$ is used. Upon substitution of Eqn.
(\ref{UZES}) into Eqn. (\ref{UEQS}), the subsequent relation is
obtain as follows.
\begin{equation}\label{1.25}
\sum\limits_{k=0}^{\infty}A_jz^{k+p}\left[(k+p)(k+p-1)-Q
z+\frac{\nu^2(\ell+1)}{f'(r_{eh})}+\frac{1}{4}\right]=0~.
\end{equation}
Our next task is to compare the coefficients for equal powers of
$z$.  By setting the coefficient of the component of $z^p$  to
zero, the following quadratic equation for $p$ is obtained.
\begin{equation}
p^2-p+\frac{\nu^2(\ell+1)}{f'(r_{eh})}+\frac{1}{4}=0,
 \label{PEQ}
\end{equation}
It is straightforward to have a solution to the preceding Eqn.
(\ref{PEQ}). The solution that comes out from (\ref{PEQ}) is
provided by
\begin{equation}
p=\frac{1}{2}\pm
i\frac{\nu\sqrt{\ell+1}}{f'(r_{eh})}~.\label{EXPP}
\end{equation}
However the following recursion relation among the coefficients in
$y$ is produced by the higher order equations:
\begin{equation}
A_n=\frac{Q
A_{n-1}}{(n+p)(n+p-1)+\frac{(\ell+1)\nu^2}{(f'(r_{eh}))^2}+\frac{1}{4}}~;~\{n=1,2,3,\ldots\}~.
\end{equation}
The generalized form of the coefficient $A_n$ with an arbitrary
value of $n$  reads
\begin{equation}
A_n=\frac{Q^n
A_0}{n!\prod\limits_{k=1}^{n}\left(k+i\frac{2\nu\sqrt{\ell+1}}{f'(r_{eh})}\right)}=\frac{Q^n
A_0\Gamma\left(1+i\frac{2\nu\sqrt{\ell+1}}{f'(r_{eh})}\right)}
{n!\Gamma\left(n+1+i\frac{2\nu\sqrt{\ell+1}}{f'(r_{eh})}\right)},
\label{EXPAN}
\end{equation}
for $p=\frac{1}{2}+\frac{i\nu\sqrt{\ell+1}}{f'(r_{eh})}$. To keep
things simple, we set  $A_0=1$. The solution considered in
Eqn.(\ref{UZES}), excluding higher order terms, has the following
structure:
\begin{equation}
\begin{split}
u(z)&=z^{\frac{1}{2}+i\frac{\nu\sqrt{\ell+1}}{f'(r_{eh})}}\sum\limits_{n=0}^\infty\frac{z^nQ^n
\Gamma\left(1+i\frac{2\nu\sqrt{\ell+1}}{f'(r_{eh})}\right)}{n!\Gamma\left(n+1+i\frac{2\nu\sqrt{\ell+1}}{f'(r_{eh})}\right)}.
\end{split}
\label{HON}
\end{equation}
Consequently, the final position and time-dependent component of
the traditional scalar field solution come out with the following
mathematical expression.
\begin{equation}
\psi_\kappa(r,t)\sim e^{-i\nu
t}z^{i\frac{\nu\sqrt{\ell+1}}{f'(r_{eh})}}\sum\limits_{n=0}^\infty\frac{z^nQ^n
\Gamma\left(1+i\frac{2\nu\sqrt{\ell+1}}{f'(r_{eh})}\right)}
{n!\Gamma\left(n+1+i\frac{2\nu\sqrt{\ell+1}}{f'(r_{eh})}\right)}~.
\label{PSIK}
\end{equation}
Note that the solution (\ref{PSIK}) has explicit dependence in
both space and time. The following section will be devoted to
calculate the transition probability resulting from a virtual
transition for an atom that falls into the event horizon of a
Lorentz-violating black hole associated with the
Einstein-bumblebee background

\section{Computation of emission and absorption probabilities}
The atom trajectory is initially required in order to compute the
transition probability. By limiting our computations to
$(3+1)$-Dimensional situation, we can make the situation simpler.
Having known the velocity $\bm{v}$, and the killing vectors
$\bm{\rho}=\partial_t$, and $\bm{\sigma}=\partial_\varphi$ we can
define two conserved quantities $\frac{E}{m}$ and $\frac{L}{m}$
respectively as follows.
\begin{equation}
\frac{E}{m}=-\bm{\rho}\cdot\bm{v}=f(r)\frac{dt}{d\tau}~,~\frac{L}{m}=\bm{\sigma}\cdot
\bm{v}=r^2\frac{d\varphi}{d\tau}. \label{EBYM}
\end{equation}
Given the invariant mass $m$, $\mathcal{E}=\frac{E}{m}$ stands for
the relativistic energy per unit rest mass of the atom and
$\mathcal{L}=\frac{L}{m}$ represents the angular momentum per unit
rest mass of the atom. The following is a recast of
Eqn.(\ref{LVSCHD}) in terms of the previously mentioned conserved
quantities in $(3+1)$-Dimensional spacetime background.
\begin{eqnarray}
&&f(r)\left(\frac{dt}{d\tau}\right)^2-\frac{\ell+1}{f(r)}\left(\frac{dr}{d\tau}\right)^2
-\frac{\mathcal{L}^2}{r^2}=1\nonumber\\
&& \frac{\mathcal{E}^2}{f(r)}
-\frac{1}{f(r)}\left(\frac{dr}{d\tau}\right)^2 =\left(1+\frac{\mathcal{L}^2}{r^2}\right)\nonumber \\
&& d\tau  =-\sqrt{\ell+1}\frac{dr}{\sqrt{\mathcal{E}^2-
f(r)\left(1+\frac{\mathcal{L}^2}{r^2}\right)}} \label{ABD}
\end{eqnarray}
Integrating Eqn.(\ref{ABD}), we have
\begin{equation}
\tau(r)=-\int\sqrt{\ell+1}\frac{dr}{\sqrt{\mathcal{E}^2
-f(r)\left(1+\frac{\mathcal{L}^2}{r^2}\right)}} ~.
\end{equation}
Using the relation in Eqn.(\ref{EBYM}) the following expressions
result.
\begin{equation}
\begin{split}
\mathcal{E}&=f(r)\frac{dt}{d\tau}\\
dt&=-\frac{\mathcal{E}\sqrt{\ell+1}}{f(r)}\frac{dr}
{\sqrt{\mathcal{E}^2-f(r)\left(1+\frac{\mathcal{L}^2}{r^2}\right)}}\\
\end{split}
\label{ABCD}
\end{equation}
Eqn. (\ref{ABCD}) after integration leads to
\begin{equation}
t(r)=-\int\frac{\mathcal{E}\sqrt{\ell+1}}{f(r)}\frac{dr}{\sqrt{\mathcal{E}^2-f(r)
\left(1+\frac{\mathcal{L}^2}{r^2}\right)}}~.
\end{equation}
As a result, the proper times $\tau$  that provide the
trajectories of the atom can be expressed as follows:
\begin{equation}
\tau(r)=-\int\frac{1}{\mathcal{E}}\sqrt{\ell+1}
\frac{dr}{\sqrt{1-\frac{f(r)}{\mathcal{E}^2}\left(1+\frac{\mathcal{L}^2}{r^2}\right)}}\label{TAU}~,
\end{equation}
where $t$ has the expression
\begin{equation}
t(r)=-\int\frac{1}{\sqrt{\ell+1}}
\frac{dr}{\sqrt{1-\frac{f(r)}{\mathcal{E}^2}\left(1+\frac{\mathcal{L}^2}{r^2}\right)}}\label{TAUE}~.
\end{equation}
Using the beyond near horizon approximation and setting
$\mathcal{L}=0$, we will now compute the trajectories of the
in-falling atom  retaining the terms up to $\mathcal{O}(z)$.
Consequently, the expressions of $\tau$ and $t$  take on the
following expressions respectively.
\begin{align}
\tau&=-\frac{z}{\mathcal{E}}\sqrt{\ell+1}+ C  \label{TAUF}~,\\
t&=-\frac{\sqrt{\ell+1}}{f'(r_{eh})}\left(\ln
z+z\left(\frac{f'(r_{eh})}{2\mathcal{E}^2}
-\frac{1}{2}\left(\frac{f''(r_{eh})}{f'(r_{eh})}\right)\right)\right)+
D,
\end{align}
The two integration constants are selected as $C$ and $D$ in this
instance.  Note that the term  $\mathcal{O}(1/z)$  is retained
in the analysis, we will first compute the transition probability.
Let us now write down the  Hamiltonian of the atom-field
interaction in order to proceed toward the computation of
transition probabilities.
\begin{equation}
\hat{\mathcal{H}}_I=\hbar\mathcal{G}\left[\hat{b}_\nu\phi_\nu(t(r),r(\tau))+H.C.\right]\left[\hat{\xi}e^{-i\omega
\tau}+ H.C.\right]. \label{XAX}
\end{equation}
Here, $\mathcal{G}$ represents the atom-field coupling constant,
$\hat{b}_\nu$ denotes the photon annihilation operator, $\omega$
is the atom transition frequency, and $\hat{\xi}$ is defined as
$\left|gr\right>\left<ex\right|$, where $\left|gr\right>$ and
$\left|ex\right>$ corresponds to the ground and excited states of a
two-level atom, respectively. The emission probability associated
with Eqn.(\ref{PHI}) can now be expressed as:
\begin{eqnarray}
P^e_{tr}&=&\frac{1}{\hbar^2}\left|\int d\tau \langle 1_\nu,ex|\hat{\mathcal{H}}_I|0,gr\rangle\right|^2 \nonumber \\
&=&\mathcal{G}^2\left|\int
d\tau~\Phi_\kappa^*(r,t)~e^{i\omega\tau}\right|^2~.
\label{TRANP}
\end{eqnarray}
The counter-rotating terms in Equation (\ref{XAX}) contribute to
the transition probability in Eqn. (\ref{TRANP}).  We make a
simple adjustment in variables, $y=\frac{\mathcal{E}x}{\omega}$,
where $x,\nu\ll \omega$ \cite{PAGE0} to proceed conveniently.
Substituting $\phi_\kappa(r,t)$ from Eqn. (\ref{PSIK})
into the equation above yields
\begin{eqnarray}
P^e_{tr} &&\cong\frac{\mathcal
{G}^2}{\omega^2}\biggr|\sqrt{1+\ell}\int_0^{\infty} (1+\frac{{\cal
E}\kappa x}{\omega(1-\frac{2i\nu\sqrt{1+\ell}}{f'(r_{eh})})} +
\frac{\mathcal{E}x}{\omega}(\frac{f'(r_{eh})} {2{\cal E}^2}
)x^{-\frac{2i\nu\sqrt{1+\ell}}{f'(r_{eh})}}
\nonumber\\
&& e^{-ix[\sqrt{1+\ell}+\frac{\mathcal {E}\nu\sqrt{1+\ell}}{\omega
f'(eh)}(\frac{f'(r_{eh})}{2e^2}
-\frac{1}{2}\frac{f''(r_{eh})}{f'(r_{eh})})]} \biggr|^2~.
\label{PTRI}
\end{eqnarray}
Another change of variables is now adopted as follows
\begin{equation}
xf'(r_{eh})=\chi\frac{f'(r_{eh})}{\sqrt{1+\ell}}~. \label{XFP}
\end{equation}
We obtain the following form of the probability under the useful
substitution of Eqn. (\ref{XFP}) in Eqn.(\ref{PTRI}).
\begin{equation}
\begin{split}
P^e_{tr}&=\frac{\mathcal{G}^2}{\omega^2}\biggr|\int
d\chi\biggr[1+\frac{\mathcal{E}\chi}{\omega}\sqrt{\frac{1}{1+\ell}}\biggr[\mu+\frac{f'(r_{eh})}{2\mathcal{E}^2}
\biggr]\\& +\frac{2i\mathcal{E}\mu\nu\chi}{\omega
f'(r_{eh})}\biggr]e^{-i\chi\left(1+\frac{\nu\mathcal{E}}{\omega
f'(r_{eh})}\left(\frac{f'(r_{eh})}{2\mathcal{E}}-\frac{1}{2}\right)\right)}
\chi^{-\frac{2i\nu\sqrt{1+\ell}}{f'(r_{eh})}}\biggr|^2,
\end{split}
\label{PTRS}
\end{equation}
where
\begin{equation}
\mu= \frac{\kappa}{1+ \frac{4\nu^2(1+ \ell)}{f(r_{eh})^2}}.
\end{equation}
The ultimate form of the excitation probability is obtained by
introducing a change in variables  as provided below
\begin{equation}
\zeta=\chi \mathcal{B}, \label{DZETA}
\end{equation}
where
\begin{equation}
\mathcal{B}=1+\frac{\nu\mathcal{E}}{2\omega
f'(r_{eh})}\left(\frac{f'(r_{eh})}{\mathcal{E}}-\frac{f''(r_{eh})}{f'(r_{eh})}\right).
\label{EEPB}
\end{equation}
We find that the  emission probability  (\ref{PTRS}) can be casted
into the following form using Eqns. (\ref{DZETA}, \ref{EEPB})
\begin{equation}
P^e_{tr} =\frac{\mathcal{G}^2}{B^2\omega^2}\biggr|\int_0^\infty
d\zeta~e^{-i\zeta}~\zeta^{-\frac{2i\nu\sqrt{1+\ell}}{f'(r_{eh})}}\biggr[1+\frac{\mathcal{E}\zeta}
{\mathcal{B}\omega
f'(r_{eh})}\left[\mu+\frac{f'(r_{eh})}{2\mathcal{E}^2}+\frac{1}{4}
\right]
+\frac{2i\mathcal{E}\mu\nu\zeta\sqrt{1+\ell}}{\mathcal{B}\omega
f'(r_{eh})}\biggr]\biggr|^2~. \label{PTRF}
\end{equation}
We will now employ another quantity provided by
\begin{equation}
\gamma=\mu+\frac{f'(r_{eh})}{2\mathcal{E}^2}.
\end{equation}
The inverse of $\mathcal{B}$ is expressed as follows:
\begin{equation}
\frac{1}{\mathcal{B}} \approx 1 - \frac{\nu}{2\omega} \left( 1 - \frac{f''(r_{eh})}{f'(r_{eh})^2} \, \mathcal{E} \right)\label{INV1}
\end{equation}
since we
include terms up to $\mathcal{O}(\frac{\nu}{\omega})$ keeping it in mind that
 $ \nu << \omega $. Consequently, the expression of
$\frac{1}{\mathcal{B}^2}$, up to the order 
$\mathcal{O}(\frac{\nu}{\omega})$ stands
\begin{equation}
\frac{1}{\mathcal{B}^2} \approx 1 - \frac{\nu}{\omega} \left( 1 - \frac{f''(r_{eh})}{f'(r_{eh})^2} \, \mathcal{E} \right).\label{IVB2}
\end{equation}
with the same consideration  $ \nu << \omega $.
The  emission probability in Eqn.(\ref{PTRF}) now acquires the
follows simplified expression.
\begin{equation}
\begin{split}
P^e_{tr}&=\frac{\mathcal{G}^2}{\omega^2\mathcal{B}^2}\biggr|\int_0^\infty
d\zeta\biggr[1+\frac{\mathcal{E}\zeta\gamma}{\mathcal{B}\omega}\sqrt{\frac{1}{1+\ell}}
+\frac{2i\mathcal{E}\mu\nu\zeta}{\mathcal{B}\omega
f'(r_{eh})}\biggr]\\&\times
e^{-i\zeta}\zeta^{-\frac{2i\nu\sqrt{1+\ell}}{f'(r_{eh})}}\biggr|^2.
\end{split}
\label{PTRE}
\end{equation}
Thus the emission probability in Eqn. (\ref{PTRE}), after a few
steps of algebra, acquires the following  form  with the  desired
Planck like factor.
\begin{equation}
P^e_{tr}\cong\frac{4\pi\mathcal{G}^2\nu\sqrt{1+\ell}}{f'(r_{eh})\mathcal{B}^2\omega^2}
\left[1-\frac{4\gamma\nu\mathcal{E}}{f'(r_{eh})\omega\mathcal{B}}
+\frac{\mathcal{E}^2\gamma^2}{(1+ \ell)\omega^2\mathcal{B}^2}
+\frac{4\nu\mathcal{E}\mu}{\mathcal{B}f'(r_{eh})\omega}\right]
\times\frac{1}{e^{\frac{4\pi\nu\sqrt{1+\ell}}{f'(r_{eh})}}-1}~.
\label{PTA}
\end{equation}
We need to compute the following factor that involves $\mathcal{B}$ 
\begin{equation}
\begin{split}
&\frac{1}{\mathcal{B}^2} \left[ 1 - \frac{4 \gamma \nu \mathcal{E}}{f'(r_{eh}) \omega \mathcal{B}} 
+ \frac{\mathcal{E}^2 \gamma^2}{(1 + \ell) \omega^2 \mathcal{B}^2} + \frac{4 \nu \mathcal{E} \mu}{\mathcal{B} f'(r_{eh}) \omega} \right]\\
&\approx 1 - \frac{\nu}{\omega} \left( 1 - \frac{f''(r_{eh})}{f'(r_{eh})^2} \mathcal{E} \right)
- \frac{4 \gamma \nu \mathcal{E}}{f'(r_{eh}) \omega}
+ \frac{\mathcal{E}^2 \gamma^2}{(1 + \ell) \omega^2}
+ \frac{4 \nu \mathcal{E} \mu}{f'(r_{eh}) \omega}\\
&\approx 1 - \frac{\nu}{\omega} (1 + 2 c^2 \mathcal{E})
- \frac{8 GM \gamma \nu \mathcal{E}}{c^2 \omega}
+ \frac{\mathcal{E}^2 \gamma^2}{(1 + \ell) \omega^2}
+ \frac{8 GM \nu \mathcal{E} \mu}{c^2 \omega}
\end{split}
\end{equation}

Thus the excitation probability is given by
\begin{equation}
P^e_{tr}\cong\frac{4\pi\mathcal{G}^2\nu\sqrt{1+\ell}}{f'(eh)\omega^2}\biggr[
1+ \frac{\nu}{\omega}\biggr(1+2\mathcal{E}) + \frac{8GM\,\mathcal{E}(\gamma-\mu)}{c^2}\biggr)\biggr]\times
\dfrac{1}{e^{\frac{4\pi\nu\sqrt{1+\ell}}{f'(r_{eh})}}-1}.\label{TRE}
\end{equation}
Analogously, the probability for  absorption,  can
be computed using the expression 
\begin{equation}
 P^a_{tr}=\frac{1}{\hbar^2}\left|\int d\tau \langle0,gr\rangle|\hat{\mathcal{H}}_I| 1_\nu,ex\right|^2 .
\end{equation}
It results in a complementary expression consistent with the principles of detailed balance
which is given by
\begin{equation}
P^a_{tr}\cong\frac{4\pi\mathcal{G}^2\nu\sqrt{1+\ell}}{f'(eh)\omega^2}\biggr[
1  + \frac{\nu}{\omega}\biggr([(1+2\mathcal{E}) + \frac{8GM\,\mathcal{E}(\gamma-\mu)}{c^2}\biggr)\biggr]\times
\dfrac{1}{1-e^{-\frac{4\pi\nu\sqrt{1+\ell}}{f'(r_{eh})}}}.\label{TRA}
\end{equation}
Note that the last factor appeared Eqn.(\ref{PTA}) is the welcome  Planck-like factor. . It should be mentioned here that  we
have ignored the $\mathcal{O}(\frac{\nu^2}{\omega^2})$ order
quantities present within parenthesis of the amplitude factor as
$\nu\ll \omega$. It is remarkable that the  excitation probability
in Eqn. (\ref{PTA}) has a Planck-like factor, which is independent
of $\kappa$ but has an explicit dependence of Lorentz volation
factor $\ell$. The following is another useful way to express the
term within the parenthesis of the  Eqn. (\ref{PTA}).

A useful and constructive comparison often invoked in the study of Hawking radiation is the correspondence between the near-horizon physics of a black hole and the radiation emitted by an accelerating mirror in Minkowski spacetime \cite{QOPT}. In particular, a uniformly accelerating mirror with constant proper acceleration $a$ produces a thermal spectrum perceived by a static observer, characterized by a definite transition probability.  In the present work, we aim to explore whether this fundamental principle continues to hold in the context of Lorentz-violating bumblebee black holes. 
The frequency $\nu$ that appears in Eqs.~(\ref{PTA},\ref{TRA}) corresponds to the frequency measured by a distant (asymptotic) observer. This asymptotic frequency, $\nu= \nu_\infty$, can be related to the locally measured frequency $\nu_o$ through the following relation:

An instructive analogy frequently employed in the analysis of Hawking radiation is the equivalence between the near-horizon physics of a black hole and the radiation produced by an accelerating mirror in Minkowski spacetime \cite{QOPT}. Specifically, a uniformly accelerating mirror with constant proper acceleration a induces a thermal spectrum detected by a static observer, characterized by the transition probability. In this work we are intended to investigate whether this fundamental principle remains valid when Lorentz violating bumblebee black hole is taken under considered
The frequency $\nu$ appearing in  Eqns. (\ref{TRE},\ref{TRA}) is
the frequency as perceived by a remote observer. Consequently,
this frequency $\nu=\nu_\infty$ can be represented in terms of the
locally observed frequency ($\nu_o$) as follows:
\begin{equation}
\nu_{\infty} = \nu_o \sqrt{f(r_{eh})} 
  \sim \nu_o \sqrt{(r-r_{eh}) f'(r_{eh})} 
  = \frac{\nu_o}{2a \sqrt{1+\ell}} f'(r_{eh}) \, .
\end{equation}
The form that results from substituting $\nu_{\infty}$ in the Planck
factors of Eqns (\ref{TRE}, \ref{TRA}) is as follows:
\begin{equation}
\dfrac{1}{e^{\frac{4\pi\nu\sqrt{1+\ell}}{f'(r_{eh})}}-1}\cong\frac{1}{e^{\frac{2\pi\nu_\mathcal{o}}{a}}-1}
\label{PLFAC}
\end{equation}      
In the vicinity of the  horizon the excitation probability  reduces to
\begin{equation}
P^e_{tr}(Q=0,NH)\cong\ =\frac{2\pi\mathcal{G}^2\nu_{o}c}{a\omega^2}\biggr[1  - \frac{\nu}{\omega}\biggr[(1+2\mathcal{E}) + \frac{8GM\,\mathcal{E}\gamma}{c^2}\
\biggr] \biggr]\times
\frac{1}{e^{\frac{2\pi\nu_{o}c}{a}}-1} \label{PTRC}
\end{equation}
when $\mu$ is neglected i.e., by setting  $\mu=0$, Consequently, spontaneous emission probability has the expression
\begin{equation}
P^e_{tr}(Q=0,NH)\cong\ =\frac{2\pi\mathcal{G}^2\nu_{o}c}{a\omega^2}\biggr[1  + \frac{\nu}{\omega}\biggr[(1+2\mathcal{E}) + \frac{8G\mathcal{E}\gamma}{c^2}\
\biggr] \biggr]\times
\frac{1}{1-e^{\frac{-2\pi\nu_{o}c}{a}}} \label{PTRC1}
\end{equation}
It is comparable to the Plank factor in the scenario of an
accelerating mirror and a fixed atom in a flat spacetime
background. This finding suggests that the gravitational force of
the black hole and the accelerated mirror have comparable effects
on the atom's field modes. The following is the excitation
probability expressed in terms of the locally measured frequency
$\nu_{o}$ and acceleration $a$.
Now recall
\begin{equation}
\mu=\frac{Q}{1+\frac{4\nu^2(1+\ell)}{f'(r_{eh})c^2}}~.
\label{EXPMU}
\end{equation}
The expression of $\mu$ in (3+1)-Dimension is as follows
\begin{equation}
\mu\cong \frac{f''(r_{eh})}{4f'(r_{eh})} +\frac{1}{r_{eh}},
\label{EXPMUA}
\end{equation}
when ${\cal L}=0$,  and the rest mass of the photon $m_0 =0$. Now
Eqn.(\ref{ABD}) leads us to have
\begin{equation}
\left(\frac{dr}{d\tau}\right)^2=\frac{g(r)}{f(r)}\mathcal{E}^2-g(r)~.
\label{EXPMUB}
\end{equation}
Note that in  the limit $r\rightarrow\infty$ both $f(r)$ and $
g(r)$ approach to unity. The  Eqn.(\ref{EXPMUB}) turns into the
following.
\begin{equation}
\left(\frac{dr}{d\tau}\right)^2=(\ell+1)(\mathcal{E}^2-1)~.
\label{EXPMUC}
\end{equation}
In the limit  $r\to\infty$, we can see from Eqn.(\ref{EXPMUC}) that
$\frac{dr}{d\tau}=0$ for $\mathcal{E}=1$. The conserved quantity
$\mathcal{E}$ represents the rest mass energy of the atom. Thus he
rest mass energy of the atom at $r\to\infty$ corresponds to
$\mathcal{E}=1$ even in the presence of the Lorentz violation
factor $\ell$. We will therefore examine the scenario where
$\mathcal{E}=1$. The emission probability in Eqn.(\ref{PTRC})
acquires the  following form in this situation with the use
implementation of  Eqn.(\ref{EXPMUA}).

It is noteworthy that the coefficient in front of the excitation
probability undergoes a change, which is dependent on the
acceleration term.  However, the Planck factor maintains its
familiar form, So, there is no violation of the equivalence principle
in spite of the violation of conformal symmetry 

The breakdown of conformal symmetry is anticipated to emerge from
considerations beyond the near-horizon approximation. It is
important to highlight that the violation of conformal symmetry does not play a substantial role in
retaining the Planck-like factor within the emission/absorption
probabilities and that too maintain the equivalence principle.

Additionally, it should be pointed out that  Eqn. (\ref{EXPMUC})
reveals that for $\mathcal{E} > 1$  $\frac{dr}{d\tau}>0 $ at
$r\to\infty$. This implies that the atom does not reach a stable
equilibrium at large radial distances $r\to \infty$.

\section{Computation of Modified HBAR entropy}
The phrase 'Horizon Brightened Acceleration Radiation Entropy'
(HBAR entropy) was initially coined in \cite{PAGE0}.  In this
section, we analyze the HBAR entropy within the context of a
Lorentz-violating Bumblebee black hole spacetime, where Lorentz
symmetry breaking plays a significant role.  Specifically, we
consider a system of two-level atoms falling into the event
horizon of the Lorentz-violating Bumblebee black hole. To compute
the HBAR entropy, we employ the density matrix formalism combined
with quantum statistical methods. Let $\omega$ is the transition
frequency of two level atom system falling into the Lorentz
violating black hole with a transition rate $\xi$. If the
microscopic change in the field density matrix is $\delta \varrho$
for one atom then the Ref.\cite{PAGE0} entails that the net
macroscopic change in the field density matrix for $\Delta N$ will
be
\begin{equation}
\Delta \rho=\sum\limits_n \delta\rho_a=\Delta N\delta\rho.
\label{DRHO}
\end{equation}
On this occasion, the  rate of fall into the event horizon denoted
by  $xi$ is define by
\begin{equation}
\frac{\Delta N}{\Delta \tau}=\xi~. \label{RFALL}
\end{equation}
Now the use of equations (\ref{DRHO},\ref{RFALL}) leads us to
write
\begin{equation}
\frac{\Delta \rho}{\Delta \tau}=\xi\delta\rho~.\label{XII}
\end{equation}
We now bring into action of the master equation for the density
matrix due to  Lindblad \cite{LIND}
\begin{equation}
\begin{split}
 \frac{d\rho}{dt}=&-\frac{\Gamma_{A}}
{2}\left(\rho a^\dagger a+b^\dagger a \rho-2a\varrho a^\dagger\right)\\
&-\frac{\Gamma_{exc}}{2}\left(\rho a a^\dagger+a a^\dagger
\rho-2a^\dagger\rho a\right),
\end{split}
 \label{MEQ}
\end{equation}
where $\Gamma_{E}$ denotes the  rate of excitation and
$\Gamma_{A}$ stands for the  rate of absorption. It is known that
$\Gamma_{E}$  and $\Gamma_{A}$ hold a relation $\Gamma_{E/A}=\xi
P_{E/A}$ where  $P_{E/A}$ is  given in Eqns.(\ref{PTRC}). If we
taking the quantum average of Eqn. (\ref{MEQ}) with respect to
some arbitrary state $|m\rangle$, we receive the following
expression
\begin{equation}
\begin{split}
\dot{\rho}_{m,m}=&-\Gamma_{A}\left(m\varrho_{m,m}-(n+1)\rho_{m+1,m+1}\right)
\\&-\Gamma_{E}\left((m+1)\varrho_{m,m}-n\varrho_{m-1,m-1}\right)~.
\end{split} \label{AME}
\end{equation}
It is useful to use the steady-state solution  to  get the
explicit expression of HBAR entropy. It allows us to make use of
the condition $\dot{\rho}_{m,m}=0$ in Eqn.(\ref{AME}).  For $n=0$
we now arrive at  the following  relation between $\rho_{1,1}$ and
$\rho_{0,0}$ as follows.
\begin{equation}
\rho_{1,1}=\frac{\Gamma_{E}}{\Gamma_{A}}\rho_{0,0}~. \label{RDEN}
\end{equation}
If we repeat this procedure  $m$ time, we land on to
\begin{equation}
\rho_{n,n}=\left(\frac{\Gamma_{E}}{\Gamma_{A}}\right)^n\rho_{0,0}~.
\label{GN}
\end{equation}
The condition $Tr(\rho)=1$ is now executed to find the explicit
expression of  $\rho_{0,0}$  in terms of $\Gamma_{E}$ and
$\Gamma_{A}$ as follows. The condition $Tr(\rho)=1$ leads us to
write
\begin{equation}
\sum\limits_m \rho_{m,m}=1,
\end{equation}
that provides
\begin{equation}
\rho_{0,0}\sum_m\left(\frac{\Gamma_{E}} {\Gamma_{A}}\right)^m=1,
\end{equation}
which ultimately results in
\begin{equation}
\rho_{0,0}= 1-\frac{\Gamma_{E}}{\Gamma_{A}}. \label{RELATION}
\end{equation}
Using $\rho_{0,0}$ from the above equation Eqn.(\ref{RELATION}),
the von-Neumann entropy for the system is computed.
\begin{equation}
\rho^{\mathcal{S}}_{n,n}=\left(\frac{\Gamma_{E}}{\Gamma_{A}}\right)^n
\left(1-\frac{\Gamma_{E}}{\Gamma_{A}}\right).
 \label{ROS}
\end{equation}
Note that for the bumblebee black hole
\begin{eqnarray}
\frac{\Gamma_{E}}{\Gamma_{A}} 
\cong& e^{-\frac{8\pi\nu GM \sqrt{1+\ell}}{c^3}}
\biggr[
1 - 2\,\frac{\nu}{\omega}\left[(1+2\mathcal{E}) + \frac{8GM\,\mathcal{E}\gamma}{c^2}\right].
\biggr], \label{VONN}
\end{eqnarray}
where $r_{eh}=\frac{2G\mathcal{M}}{c^2}$. To obtain Eqn.
(\ref{VONN}), 
We have introduced the  assumptions $f'(r_{ec}),f''(r_{ec})$ and $\nu\ll\omega$,
The von-Neumann entropy and its rate of change for the system are given by
\cite{PAGE0}
\begin{align}
S_\rho &=-k_B\sum\limits_{n,\nu}\rho_{n,n}\ln\rho_{n,n}~,\\
\dot{S}_\rho&=-k_B\sum\limits_{n,\nu}\dot{\rho}_{n,n}\ln\rho_{n,n}\cong
-k_B\sum\limits_{n,\nu}\dot{\rho}_{n,n}\ln\rho^{\mathcal{S}}_{n,n}
\label{VON}
\end{align}
where in case of the rate of change, we have replaced $\rho_{n,n}$
by the steady state solution $\rho_{n,n}^{\mathcal{S}}$. Using the
form of $\rho_{n,n}^{\mathcal{S}}$ from Eqns.(\ref{VONN},
\ref{VON}), we obtain the analytical form of $S_\rho$ as follows
\begin{eqnarray}
\dot{S}_\rho &\cong& -k_B\sum\limits_{n,\nu}n\dot{\rho}_{n,n}
\biggr[-\frac{8\pi\nu G
\mathcal{M}\sqrt{\ell+1}}{c^3}+\ln\biggr(1-2 \frac{\nu}{\omega} \left(1 + 2 \mathcal{E} \right) - \frac{16G M\gamma \nu \mathcal{E}}{c^2\omega}\biggr)\biggr]\nonumber\\
&\cong & \biggr[\frac{8\pi G\mathcal{M}\sqrt{\ell+1}}{c^3}+ \frac{2}{\omega} \left(1 + 2 \mathcal{E} \right)  +\frac{16G M\gamma \mathcal{E}}{c^2\omega}
\biggr]k_B\sum\limits_{\nu}\dot{\bar{n}}_\nu\nu,
 \label{SDOT}
\end{eqnarray}
where $\bar{n}_\nu$ is the photon flux generated from two-level
atoms in the vicinity of the event horizon of the black hole.
\begin{eqnarray}
\dot{S}_\rho(\mathcal{E}=1 )\cong  \biggr[\frac{8\pi G\mathcal{M}\sqrt{\ell+1}}{c^3}+ \frac{6}{\omega}   +\frac{16G M\gamma }{c^2\omega}
\biggr]k_B\sum\limits_{\nu}\dot{\bar{n}}_\nu\nu
\end{eqnarray}
The net loss of energy due to emitted photons is given by
\begin{equation}
\hbar\sum\limits_\nu \dot{\bar{n}}_\nu\nu=\dot{m}_pc^2.
\end{equation}
The surface area of the black hole under consideration is
\begin{equation}
A_{LVB}=4\pi r^2_{eh} = \cong\frac{16\pi G^2\mathcal{M}^2}{c^4}.
\label{AREA}
\end{equation}
The time derivative of $A_{LVB}$ yields
\begin{equation}
\dot{A}_{LVB}=\frac{32\pi G^2\mathcal{M}\dot{\mathcal{M}}}{c^4}.
\end{equation}
Consequently, the rate of change of mass can be expressed as
\begin{equation}
\dot{\mathcal{M}}=\dot{m}_{LVB}+\dot{m}_{atom}. \end{equation}
Here, $M$ stands for mass of the black hole, and  $\dot{m}_{ep}$
is the rate of change of the rest mass of the black hole is due to
emitting photons \cite{PAGE0}. We now define the rate of change in
the area of the black hole due to the emitted photons  by the
expression
\begin{equation}
\dot{A}_{LVB}=\frac{32\pi G^2\mathcal{M}\dot{m}_{ep}}{c^4}.
\label{RCA}
\end{equation}
From the above computation, we can draw an inference that $A_{ep}$
is the area of the black hole when no atom is falling into the
black hole. So, $A_{atom}$ can be considered as vanishingly small.
This result can be interpreted in the following way. The atom
emits radiation prior to entering the event horizon. Thus, the
black hole entropy associated with HBAR radiation from an atom and
that associated with an atom can be segregated. Hence, it is
acceptable that $A_{ep}$ is equal to the area of the entire black
hole when no atoms are falling into the black hole. So the
expression of $\dot{S}_\rho$ in terms of $A_{ep}$ in closed form
reads
\begin{equation}
\dot{S}_\rho =\frac{d}{dt}\biggr[F(A_{ep})
\biggr]. \label{DEN}
\end{equation}
where 
\begin{equation}
F(A_{ep})=\frac{k_Bc^3}{4\hbar
G}A_{ep}\sqrt{\ell+1}  + \frac{3k_Bc^4}{2\hbar\omega\sqrt{\pi}G}A^{\frac{1}{2}}_{ep} +\frac{\gamma  k_B M^2}{2\hbar \omega c^2}\ln A_{LVB}
\end{equation}

Equation (\ref{DEN}) explicitly reveals the dependence on the
Lorentz violation factor, marking a novel finding of our study.
Furthermore, we observe that the coefficient of the second term in
Eqn. (\ref{SDOT}) is modified in Equation (\ref{DEN}), with the
emergence of two additional $\hbar$-dependent terms. These terms
are likely a consequence of extending the analysis beyond the
near-horizon approximation, as discussed in the prior works
\cite{BAKE1, BAKE2, HAW1, HAW2, PK}. The emergence of logarithmic corrections is particularly noteworthy,
as they arise in a wide range of scenarios where quantum effects become significant. Such corrections 
are not confined to a single framework; rather, they appear across diverse contexts—from black hole entropy and quantum field 
theory in curved spacetime to the corrections in thermodynamic quantities. Their prevalence suggests that logarithmic 
terms encode a universal imprint of quantum fluctuations on classical backgrounds. In the present setting, the appearance 
of these corrections provides valuable insight into the interplay between quantum effects and fundamental principles, 
such as the equivalence principle, and highlights the need for a deeper investigation into their physical origin and implications.

\section{Conclusion}
In this study, we investigated the interaction of a two-level atom with a Schwarzschild-like black hole 
in a Lorentz-violating background, specifically modeled within the framework of bumblebee gravity. To isolate the system from the 
influence of Hawking radiation, we assumed that the black hole is enclosed by a perfectly reflecting mirror, ensuring that the initial 
field configuration appears vacuum-like to an external observer.

Our analysis centered on two main aspects: computation of the excitation probability of the atom-field system and the evolution of the HBAR entropy.
We found that the excitation probability exhibits a modified Planck-like factor, signaling the emission of real photons. Notably, the familiar Planck factor 
remain unaltered due to the Lorentz-violating effects encoded in the modified frequency spectrum, 
highlighting the influence of quantum gravity corrections introduced by the bumblebee model within the amplitude part.

Interestingly, our results indicate that the Equivalence Principle remains preserved despite the presence of Lorentz-violating factors and 
consequent 
braking of apparent conformal symmetry
This suggests that Lorentz violation, as introduced by bumblebee gravity, does not inherently lead to a breakdown of equivalence principle.
Additionally, our findings resonate with earlier studies that examined corrections arising from various modified spacetime geometries.

The study also reveals an intricate interplay between Lorentz violation, conformal symmetry breaking, and the resulting radiative processes. 
While neither Lorentz violation nor conformal symmetry breaking individually violates the Equivalence Principle, 
both significantly influence the radiation amplitude and the thermodynamic behavior of the system.

In summary, our work demonstrates that Lorentz-violating effects which is manifested through bumblebee gravity, can lead to meaningful modifications 
in black hole thermodynamics, particularly in the context of acceleration radiation and entropy evolution, 
all while preserving the fundamental equivalence principles of  theory of gravity and the welcome logarithmic correction evolves.

\end{document}